\documentclass[reprint, amsmath, amssymb, aps]{revtex4-1}

\usepackage[pdftex]{graphicx}
\usepackage{wasysym}
\usepackage{amsfonts}
\usepackage{ifsym}
\usepackage{ulem}

\newcommand{\nn}{\nonumber}

\newcommand{\COMMENT}[1]{}

\newcommand{\neqa}{\nonumber\end{eqnarray}}
\newcommand{\la}[1]{\label{#1}}

\newcommand{\<}{{\langle}}
\renewcommand{\>}{{\rangle}}

\newcommand{\re}{\relax{\rm I\kern-.18em R}}

\def\su2{{SU(2)}}

\def\[{\left[}
\def\]{\right]}

\def\({\left(}
\def\){\right)}
\def\[{\left[}
\def\]{\right]}

\def\<{\langle}
\def\>{\rangle}

\def\i2{\frac{i}{2}}

\def\2F1{\,_2{\rm F}_1}

\usepackage[usenames,dvipsnames]{xcolor}
\usepackage{color}
\usepackage{graphicx}
\usepackage{dcolumn}
\usepackage{bm}
\usepackage{hyperref}

\newcommand{\beq}{\begin{equation}}
\newcommand{\eeq}{\end{equation}}
\newcommand{\beqq}{\begin{equation*}}
\newcommand{\eeqq}{\end{equation*}}
\newcommand\beqa{\begin{eqnarray}}
\newcommand\eeqa{\end{eqnarray}}
\newcommand\beqaa{\begin{eqnarray*}}
\newcommand\eeqaa{\end{eqnarray*}}
\newcommand\bea{\begin{array}}
\newcommand\eea{\end{array}}

\begin{document}

\title{Hexagonal Wilson Loops in Planar $\mathcal{N}=4$ SYM Theory at Finite Coupling}

\author{Benjamin Basso$^{\displaystyle\hexagon}$, Amit Sever$^{{\displaystyle\Box}}$ and Pedro Vieira$^{\displaystyle\pentagon}$}

\affiliation{
\vspace{5mm}
$^{\displaystyle\hexagon}$Laboratoire de Physique Th\'eorique, \'Ecole Normale Sup\'erieure, Paris 75005, France\\
$^{\displaystyle\Box}$School of Physics and Astronomy, Tel Aviv University, Ramat Aviv 69978, Israel\\
$^{\displaystyle\pentagon}$Perimeter Institute for Theoretical Physics,
Waterloo, Ontario N2L 2Y5, Canada
}

\begin{abstract}
\vspace{1mm}
We report on the complete OPE series for the 6-gluon MHV and NMHV amplitudes in planar $\mathcal{N}=4$ SYM theory. Namely, we provide a finite coupling prediction for all the terms in the expansion of these amplitudes around the collinear limit. These furnish a non-perturbative representation of the full amplitudes.

\end{abstract}

\maketitle

In~\cite{short}, building on \cite{OPEpaper}, the so-called POPE (Pentagon Operator Product Expansion) program was put forward. Its goal is to put to full use the integrability of the chromodynamic flux tube of planar $\mathcal{N}=4$ SYM theory to study scattering amplitudes and polygonal Wilson loops at any value of the {'t Hooft} coupling {$g = \sqrt{\lambda}/4\pi$}. 

Since then, through a series of works \cite{data,2pt,Andrei1,O6paper,gluons,Andrei2,Andrei3,shortSuper,Andrei4,heptagonPaper}, a simple physical and mathematical structure underpinning these objects has emerged. In this short letter, we collect all known information about hexagonal loops -- i.e. 6-gluon scattering amplitudes -- and present their complete non-perturbative expression as a POPE expansion. 

Within the POPE formalism, the {renormalized hexagonal} Wilson loops \cite{OPEpaper,short} are given by an infinite sum 
\beq
\centering\def\svgwidth{7cm}
\begingroup%
  \makeatletter%
  \providecommand\color[2][]{%
    \errmessage{(Inkscape) Color is used for the text in Inkscape, but the package 'color.sty' is not loaded}%
    \renewcommand\color[2][]{}%
  }%
  \providecommand\transparent[1]{%
    \errmessage{(Inkscape) Transparency is used (non-zero) for the text in Inkscape, but the package 'transparent.sty' is not loaded}%
    \renewcommand\transparent[1]{}%
  }%
  \providecommand\rotatebox[2]{#2}%
  \ifx\svgwidth\undefined%
    \setlength{\unitlength}{820.24375bp}%
    \ifx\svgscale\undefined%
      \relax%
    \else%
      \setlength{\unitlength}{\unitlength * \real{\svgscale}}%
    \fi%
  \else%
    \setlength{\unitlength}{\svgwidth}%
  \fi%
  \global\let\svgwidth\undefined%
  \global\let\svgscale\undefined%
  \makeatother%
  \begin{picture}(1,0.2927162)%
    \put(0,0){\includegraphics[width=\unitlength]{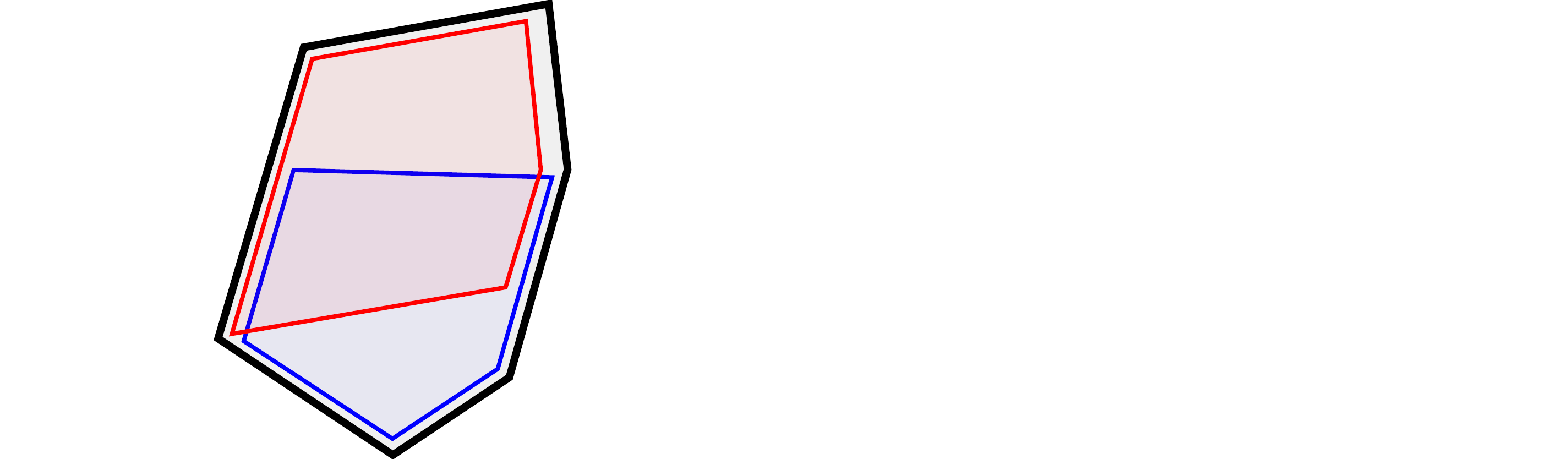}}%
    \put(-0.00246399,0.14137296){\color[rgb]{0,0,0}\makebox(0,0)[lb]{\smash{$\displaystyle{\mathcal{W}_6=}$}}}%
    \put(0.3967976,0.14074378){\color[rgb]{0,0,0}\makebox(0,0)[lb]{\smash{$\displaystyle{={\sum\limits_{n} \frac{1}{S_n} \int
\frac{d u_1\dots du_n}{(2\pi)^n}}
\,\Pi(\{u_i\})}$}}}%
  \end{picture}%
\endgroup%
\la{sum} 
\eeq
{over all possible $n$-particle states of the colour flux tube, integrated over their $n$-dimensional phase space,} and with $S_{n}$ a symmetry factor for identical particles; it is equal to $n_{1}!n_{2}!\ldots n_{k}!$ with $n=n_{1}+\ldots n_{k}$ and $n_{j}$ the number of identical particles of a given kind. {The POPE integrand in~(\ref{sum}) stands for the probability that the corresponding state gets successively produced and annihilated on, respectively, the bottom and top pentagons inside the hexagon. It} can be conveniently decomposed as a product of {three} simpler building blocks as
\beq\la{integrand}
\Pi =  \Pi_\text{dyn} \times\Pi_{\textrm{FF}}\times \Pi_{\textrm{mat}}\, .
\eeq
{The product of the first two terms was spelled out explicitly in~\cite{heptagonPaper}. The last factor, the so-called matrix part, deals with the $SU(4)$ R-symmetry degrees of freedom of the excitations,} and was partially unveiled in~\cite{O6paper} for scalar excitations in the hexagon MHV Wilson loop. Below we quote this last missing factor for any state on any hexagon.  The derivation of this matrix part factor and its generalization to any polygon will be given in~\cite{Frank}. 

The \textit{dynamical part} neatly factorizes as
\beq
\Pi_\text{dyn}=\prod\limits_i\mu(u_i)\,e^{-E(u_i)\tau+ip(u_i)\sigma+im_i\phi} \,{\prod_{i<j}\frac{1}{|P(u_{i}|u_{j})|^2}}\, , \nn
\eeq
where $\tau$, $\sigma$ and $\phi$ are the thee independent conformal cross ratios we denote in the OPE as time, space and angle, see \cite{short}. The hexagon geometry enters only in these exponentials. Most importantly, we have the pentagon transition link $P(u_{i}|u_{j})$ between excitation $i$ and excitation $j$ and their corresponding flux tube measures $\mu(u_i)$.

Next we have the {\textit{form factor part}} given once more by a simple factorized product 
\beq\la{ffpart}
\Pi_{\textrm{FF}} =g^{\frac{r_b(r_b-4)}{8}+\frac{r_t(r_t-4)}{8}}\times\prod_{i} h(u_i)^{r_t-r_b}\, ,\nn
\eeq
where $r_b$ ($r_t$) {$=0, 1, \ldots, 4$} is the R-charge carried by the bottom (top) pentagon \cite{shortSuper,heptagonPaper}. {For NMHV hexagons we have $r_t=4-r_b$, while for MHV and N$^2$MHV amplitudes we have $r_b=r_t=0$ and $r_b=r_t=4$, respectively. Therefore this factor is only needed to describe (any of the five independent) NMHV hexagons.}

Note that $P$, $\mu$, $h$ and the propagation charges implicitly depend on the flavour of the excitation; e.g.~if excitation $i$ is a fermion $\psi$ and excitation $j$ a scalar $\phi$, then in the notation of~\cite{shortSuper} we have $P(u_{i}|u_{j}) = P_{\psi\phi}(u_{i}|u_{j})$, $\mu(u_i)=\mu_\psi(u_i)$, $h(u_i)=h_\psi(u_i)$, $E(u_i)=E_\psi(u_i)$ etc. All form factors, transitions and measures are summarized in~\cite{heptagonPaper} while the dispersion relations can be found in~\cite{2pt}. For scalar and gluon excitations the contour of the integration in (\ref{sum}) is over the real axis. For fermions it is slightly more involved. As described in \cite{data}, for these excitations it is often convenient to divide the integration contour into so-called large and small fermion contours. 

Finally we have the \textit{matrix part} which encodes the $SU(4)$ R-symmetry structure of the theory. It comes about from the contraction of R-charge indices between the bottom and top pentagons. The matrix part is a coupling independent rational function of (differences of) the rapidities $\{u_i\}$. For generic number of excitations it does not exhibit any obvious factorization, in contrast with the two contributions described thus far. As an example, consider a three-scalar contribution in an NMHV hexagon with two units of R-charge at the bottom and similarly at the top. For such state the matrix part is given by 
\beq
\!\!\! \Pi_\text{mat}=\frac{ \big(7+\sum\limits_{i} s_i^2-\sum\limits_{i<j}s_i s_j\big) \big(12+\sum\limits_{i}s_i^2-\sum\limits_{i<j}s_i s_j\big) }{\frac{1}{6i}\prod\limits_{i<j}\left(\left(s_i-s_j\right){}^2+1\right)
   \left(\left(s_i-s_j\right){}^2+4\right) } \la{example3phi}
\eeq
where $\{s_{1},s_2,s_3\}$ are the rapidities of the three scalars. The matrix part for states with more scalars or fermions typically lead to gigantic rational functions.
Nonetheless, viewed from an appropriate angle, a neat factorization is actually there \textit{even} for the matrix part!   

\begin{figure}[t]
\centering
\def\svgwidth{7cm}
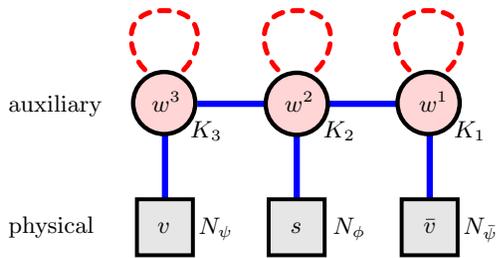
\caption{The pattern of auxiliary rapidities entering the matrix part has a simple group theoretical interpretation. The three sets of rapidities $w^{1,2,3}$ can be identified with the three nodes of the $SU(4)$ Dynkin diagram. The occupation numbers $K_{1,2,3}$ are fixed such that the overall state with $N_\phi$ scalars, $N_\psi$ fermions and $N_{\bar\psi}$ anti-fermions belongs to the $SU(4)$ representation $\{{\bf 1},{\bf\bar 4},{\bf 6},{\bf4},{\bf 1}\}$ for $r_b=\{0,1,2,3,4\}$ respectively. The cartoon depicted here is related to (\ref{matrixpart}) by identifying the solid lines with $f$'s and the dashed lines with $g$'s. }\label{auxiliaryroots}
\end{figure}

Indeed, it so happens that this matrix part itself admits an integral representation over auxiliary rapidities with a very simple and totally factorized integrand. To describe it one must split the rapidities $\{u_j\}$ into those carried by the scalars, fermions and antifermions, which we denote as $\{s_i\}, \{v_i\}$ and $\{\bar v_i\}$, respectively.
The remaining rapidities  -- those of the gluons -- do not enter the matrix part since gluons do not carry R-charge. Then
\beqa
&&\!\!\!\!\Pi_\text{mat}={1\over K_1!K_2!K_3!}\int\prod\limits_{i=1}^{K_1}{dw_i^{1}\over2\pi}\int\prod\limits_{i=1}^{K_2}{dw_i^{2}\over2\pi}\int\prod\limits_{i=1}^{K_3}{dw_i^{3}\over2\pi} \nn\\
&&\!\!\!\!\quad\times{g({\bf w}^1)g({\bf w}^2)g({\bf w}^3)\over f({\bf w}^1,{\bf w}^2)f({\bf w}^2,{\bf w}^3)f({\bf w}^1,{\bf v})f({\bf w}^2,{\bf s})f({\bf w}^3,{\bf\bar v})}\la{matrixpart} \nn \,,
\eeqa
where all integrals are over the real axis and where we use the short-hand notation 
\beq
g({\bf w})=\prod\limits_{i<j}g(w_i-w_j)\ ,\quad f({\bf w},{\bf v})=\prod\limits_{i,j}f(w_i-v_j) \nn \,,
\eeq
with $f(x)=x^2+1/4$ {and} $g(x)=x^2(x^2+1)$ to describe the fully factorized matrix part integrand, also depicted in figure \ref{auxiliaryroots}.  The numbers of auxiliary roots 
$K_{1,2,3}$ are given by the solution to 
\beqa
&&N_\psi - 2 K_1 + {\color{white}2}K_2{ \color{white} -2K_3}\,\,\,= \delta_{r_b,3} \,, \nn \\
&&N_{\phi} \,+ {\color{white}2}K_1- 2 K_2 + {\color{white}2}K_3 = \delta_{r_b,2}\,, \nn \\
&&N_{\bar\psi} {\color{white}-2K_3\,}\,\,+{\color{white}2}K_2-\, 2 K_3 = \delta_{r_b,1}\,.  \nn
\eeqa
All integrals in the matrix part representation can be straightforwardly performed by residues. For an NMHV hexagon with~$r_b=2$, $N_{\phi}=3$ and $N_{\psi} = N_{\bar \psi}=0$, for instance, we readily reproduce (\ref{example3phi}) in this way.

This is it. All these ingredients can now be straightforwardly plugged in (\ref{sum}) to yield a fully non-perturbative representation of any six-point scattering amplitudes in planar $\mathcal{N}=4$ SYM theory at any value of the coupling. 

As usual with bootstrap based approaches, our result lies somewhere between a physics conjecture and a mathematical theorem. Any efforts to rigorously establish (\ref{sum}) or to yield further evidence for it would be {most valuable.} 

In the meantime, it is a fascinating problem to investigate the POPE sum in detail and look for physically interesting regimes where it simplifies. 

One such limit is weak coupling. At weak coupling almost all integrals can be trivially done. More precisely, all fermion integrals over the so-called small fermion domain can be performed straightforwardly leaving one with the integration over the rapidities of the others excitations -- be them large fermions, scalars or gluons. The number of such integrations grows very slowly in perturbation theory. For MHV say, at~$l$ loops, we have at most~$\sqrt{2l}$ non-trivial integrations to perform! This is extremely economical as compared to any perturbative expansion and is one of the reasons why the POPE approach is so convenient as a boundary data provider for the so-called hexagon bootstrap program \cite{LanceEtAl1,LanceEtAl2,LanceEtAl3,LanceEtAl4,LanceEtAl5}. {Reproducing the full kinematical dependence of the $l$-loop amplitude, away from the collinear regime, still requires summing over infinite families of flux tube excitations (arising mostly from integrating out any number of small fermions as discussed above). This calls for the development of efficient methods for both performing the OPE integrals and re-summing them, as tackled recently in \cite{georgiosjames,georgios} for the so called gluonic double scaling limit \cite{gluons}.} 

Another very interesting regime is strong coupling. Equipped with the POPE representation (\ref{sum}) one might want to derive the full strong coupling classical result~\cite{AM,bubble,Ysystem}, and address the question of computing sub-leading quantum corrections, {following analyses of~\cite{strongC1,strongC2,O6paper,strongC3,strongC4} for amplitudes and similar problems.}

It is an equally fascinating problem to try to access other physically relevant kinematical regimes starting from the OPE sum. {One such regime is the Regge limit \cite{regge1,Simon,regge2,regge3} tackled from the POPE lens in \cite{BFKL,Yasuyuki,BLP}.}
{It is also very interesting to establish a relation to the Q-bar approach \cite{Qbar1,Qbar2} as initiated in \cite{Andrei5}.}

Finally, we look forward to Monte Carlo simulating the flux tube gas (\ref{sum}) on a computer and plot a six gluon amplitude all the way from weak to strong coupling~\cite{MattEtAl}.

Rare are the examples of CFT correlators for which the complete OPE series was found. {Excitingly, thanks to integrablity, this turned out to be possible here.}

\section*{Acknowledgements} 
We are obliged to J.~Caetano, L.~Cordova, F.~Coronado, V.~Kazakov and M.~von Hippel for illuminating discussions and specially to L.~Dixon for invaluable discussions and data exchanges over the past years.
Research at the Perimeter Institute is supported in part by the Government of Canada through NSERC and by the Province of Ontario through MRI. A.S. has been supported by the I-CORE Program of the Planning and Budgeting Committee, The Israel Science Foundation (grant No. 1937/12) and the EU-FP7 Marie Curie, CIG fellowship.

\end{document}